\documentclass[runningheads]{llncs}

\usepackage[utf8]{inputenc}
\usepackage{url}
\usepackage{amsmath}
\usepackage{mathbbol}
\usepackage{mathtools}
\usepackage{hyperref}
\usepackage{todonotes}
\usepackage{appendix}
\usepackage[strict]{changepage}

\usepackage{multicol}

\usepackage{listings}

\usepackage{tikz}

\usepackage{bbm}

\usepackage{mathpartir}

\usepackage{makecell}

\spnewtheorem{nidefinition}[definition]{Definition}{}{}
\spnewtheorem{nilemma}[lemma]{Lemma}{\bfseries}{}
\spnewtheorem{nitheorem}[theorem]{Theorem}{\bfseries}{}

\AtBeginDocument{%

}

\newcommand{\lambdacalc}{$\lambda$-calculus}
\newcommand{\picalc}{$\pi$-calculus}

\newcommand{\stacked}[1]{\mprset{flushleft} \inferrule*{}{#1}}
\newcommand{\datatype}[2]{{\mprset{fraction={===}} \inferrule{#1}{#2}}}

\newcommand{\type}[1]{\textcolor{blue}{\operatorname{#1}}}
\newcommand{\constr}[1]{\textcolor{orange}{\operatorname{#1}}}
\newcommand{\func}[1]{\textcolor{teal}{\operatorname{#1}}}

\newcommand{\PO}{\constr{\mathbb{0}}}
\newcommand{\comp}[2]{#1 \, \constr{\parallel} \, #2}
\newcommand{\new}{\constr{\boldsymbol{\nu}} \,}
\newcommand{\send}[2]{#1 \, \constr{\langle} \, #2 \,\constr{\rangle} \,}
\newcommand{\recv}[2]{#1 \, \constr{\mathbb{(}} \, #2 \, \constr{\mathbb{)}} \,}
\newcommand{\suc}{\constr{\scriptstyle 1+}}
\newcommand{\unit}{\constr{\mathbbm{1}}}

\newcommand{\channel}[2]{\constr{C[} \, #1 \, \constr{;} \, #2 \, \constr{]}}
\newcommand{\comma}{\, \constr{,} \,}

\newcommand{\subst}[3]{#1 \, \func{[} \, #3 \, \func{\mapsto} \,#2 \, \func{]}}
\newcommand{\op}[3]{#1 \, \func{\coloneqq} \, #2 \, \func{\cdot} \, #3}
\newcommand{\opsquared}[3]{#1 \, \func{\coloneqq} \, #2 \, \func{\cdot^2} \, #3}
\newcommand{\opctx}[3]{#1 \, \func{\coloneqq} \, #2 \, \func{\otimes} \, #3}
\newcommand{\zero}{\func{\cdot-0}}
\newcommand{\one}{\func{\cdot-1}}
\newcommand{\li}{\func{\ell_i}}
\newcommand{\lo}{\func{\ell_o}}
\newcommand{\lz}{\func{\ell_{\emptyset}}}
\newcommand{\lio}{\func{\ell_{\#}}}

\newcommand{\Set}{\type{SET}}
\newcommand{\reduce}[1]{\, \type{\longrightarrow}_{#1} \,}
\newcommand{\types}[4]{#1 \, \type{;} \, #2 \, \type{\vdash} \, #3 \, \type{\triangleright} \, #4}
\newcommand{\contains}[6]{#1 \, \type{;} \, #2 \, \type{\ni}_{#3} \, #4 \, \type{;} \, #5 \, \type{\triangleright} \, #6}
\newcommand{\containsusage}[4]{#1 \, \type{\ni}_{#2} \, #3 \, \type{\triangleright} \, #4}
\newcommand{\Var}{\type{VAR}}
\newcommand{\Process}{\type{PROCESS}}
\newcommand{\Raw}{\type{RAW}}
\newcommand{\Name}{\type{NAME}}
\newcommand{\Names}{\type{NAMES}}

\newcommand{\WellScoped}{\type{WELLSCOPED}}
\newcommand{\Unused}{\type{UNUSED}}
\newcommand{\PreCtx}{\type{PRECTX}}
\newcommand{\Ctx}{\type{CTX}}
\newcommand{\Type}{\type{TYPE}\,}
\newcommand{\Idx}{\type{IDX}\,}
\newcommand{\Idxs}{\type{IDXS}}
\newcommand{\Usage}{\type{USAGE}}
\newcommand{\N}{\type{\mathbb{N}}}
\newcommand{\Channel}{\type{CHANNEL}}

\newcommand{\Algebra}{\type{ALGEBRA}}
\newcommand{\eq}{\, \type{\simeq} \,}
\newcommand{\eqeq}{\, \type{\cong} \,}

\title{$\pi$ with leftovers: \\ a mechanisation in Agda
\thanks{This work is supported by the EU HORIZON 2020 MSCA RISE project 778233 ``Behavioural Application Program Interfaces'' (BehAPI).}}
\titlerunning{$\pi$ with leftovers: a mechanisation in Agda}
\author{
  Uma Zalakain\inst{1}\orcidID{0000-0002-3268-9338}
  \and
  Ornela Dardha\inst{2}\orcidID{0000-0001-9927-7875}
}
\authorrunning{U. Zalakain \and O. Dardha}

\institute{
  University of Glasgow, Scotland \ \ \email{u.zalakain.1@research.gla.ac.uk}
  \and
  University of Glasgow, Scotland \ \ \email{ornela.dardha@glasgow.ac.uk}
}

\begin{document}

\maketitle

\begin{abstract}
  Linear type systems need to keep track of how programs use their resources.
  The standard approach is to use \emph{context splits} specifying how resources are (disjointly) split across subterms.
  In this approach, context splits redundantly echo information which is already present within subterms.  An alternative approach is to use \emph{leftover typing} \cite{Mackie,Allais2018a}, where in addition to the usual (input) usage context, typing judgments have also an output usage context: the leftovers.
  In this approach, the leftovers of one typing derivation are fed as input to the next, threading through linear resources while avoiding context splits.
  We use leftover typing to define a type system for a {resource-aware \picalc{}} \cite{MilnerPW92,Milner99}, a process algebra used to  model concurrent systems.
  Our type system is parametrised over a set of \emph{usage algebras} \cite{JungSSSTBD15,TuronTABD13} that are general enough to encompass \emph{shared types} (free to reuse and discard), \emph{graded types} (use exactly $n$ number of times) and \emph{linear types} (use exactly once).
   Linear types are important in the \picalc{}: they ensure privacy and safety of communication and avoid race conditions, while graded and shared types allow for more flexible programming.
  We provide a framing theorem for our type system, generalise the weakening and strengthening theorems to include linear types, and prove subject reduction.
  Our formalisation is fully mechanised in about 1850 lines of Agda \cite{Zalakain2020Agda}.

\end{abstract}

\section{Introduction}

The \picalc{} \cite{MilnerPW92,Milner99} is a computational model for communication and concurrency that boils concurrent processing down to the sending and receiving of data over communication channels.
Notably, it features channel mobility: channels themselves are first class values and can be sent and received.
Kobayashi et al. \cite{KPT96} introduced a typed version of the \picalc{} with linear channel types, where channels must be used \emph{exactly} once.
Linearity in the \picalc{} guarantees privacy and safety of communication and avoids race conditions.

More broadly, linearity allows for resource-aware programming and more \emph{efficient} implementations \cite{Wadler90}, and it  inspired unique types (as in Clean \cite{BarendsenS96}), and ownership types (as in Rust \cite{MatsakisK14}).
A linear type system must keep track of what resources are used in which parts of the program, and guarantee that they are \emph{neither duplicated nor discarded}.
To do so, the standard approach is to use context splits: typing rules for terms with multiple subterms add an extra side condition specifying what resources to allocate to each of the subterms.
The typing derivations for the subterms must then use the entirety of their allocated resources.
A key observation here is that each subterm already \emph{knows} about the resources it needs.
\emph{Context splits contain usage information that is already present in the subterms.}
Moreover, the subterms cannot be typed until the context splits have been defined.
On top of that, using binary context splits means that typing rules with $n$ subterms require $n - 1$ context splits, which considerably clutters the type system.

An alternative approach is \emph{leftover typing}, a technique used to formulate intuitionistic linear logic \cite{Mackie} and to mechanise the linear \lambdacalc~\cite{Allais2018a}.
Leftover typing changes the shape of the typing judgments and includes a second \emph{leftover} output context that contains the resources that were left unused by the term.
As a result, typing rules \emph{thread} the resources \emph{through} subterms without needing context splits: each subterm uses the resources it needs, and leaves the rest for its siblings.
The first subterm in this chain of resources immediately knows what resources it has available.

In this paper, we use leftover typing to define for the first time a resource-aware type system for the \picalc{}, and we fully mechanise our work in Agda \cite{Zalakain2020Agda}. 
All previous work on mechanisation of linear process calculi uses context splits instead \cite{Gay2001,Goto2016a,Gay2010,Thiemann2019,Ciccone}. We will further highlight the benefits of leftover typing as opposed to context splits in contributions and the rest of the paper.

Below we present two alternative typing rules for parallel composition in the linear \picalc{}: the one on the left uses context splits, while the one on the right does not, and uses leftover typing instead:
\begin{mathpar}
  \inferrule
  {\opctx{\Gamma}{\Delta}{\Xi} \\ \Delta \; \type{\vdash} \; P \\ \Xi \; \type{\vdash} \; Q}
  {\Gamma \; \type{\vdash} \; \comp{P}{Q}}

  \inferrule
  {\Gamma \; \type{\vdash} \; P \; \type{\triangleright} \; \Delta \\ \Delta \; \type{\vdash} \; Q \; \type{\triangleright} \; \Xi}
  {\Gamma \; \type{\vdash} \; \comp{P}{Q} \; \type{\triangleright} \; \Xi}
\end{mathpar}

\paragraph*{Contributions and Structure of the Paper}
\begin{enumerate}
  \item \textbf{Leftover typing for resource-aware \picalc{}}. Our type system uses leftover typing to model the resource-aware \picalc{} (\autoref{leftover-typing}) and satisfies subject reduction (\autoref{thm:subject-reduction}).
    In addition to making context splits unnecessary, leftover typing allows for a \emph{framing} theorem (\autoref{thm:framing}) to be stated and is naturally associative, making type safety properties considerably easier to reason about (\autoref{meta-theory}).
    Thanks to leftover typing, we can now state \emph{weakening} (\autoref{thm:weakening}) and \emph{strengthening} (\autoref{thm:strengthening}) for the whole framework, not just the shared fragment. This give a uniform and complete presentation of all the meta-theory for the resource-aware \picalc{}.    
  \item \textbf{Shared, graded and linear unified \picalc{}}.
  We {generalise resource counting to a set of usage algebras that can be mixed within the same type system}.
    We do not instantiate our type system to only work with linear resources, instead we present an algebra-agnostic type system, and admit a mix of user-defined \emph{resource aware} algebras \cite{JungSSSTBD15,TuronTABD13} (\autoref{multiplicities}).
    Any \emph{partial commutative monoid} that is \emph{decidable}, \emph{deterministic}, \emph{cancellative} and has a \emph{minimal element} is a valid such algebra.
    Multiple algebras can be mixed in the type system --- usage contexts keep information about what algebra to use for each type (\autoref{contexts}).
    In particular, this allows for type systems combining linear (use exactly once), graded (exact number of $n$ times) and shared (free to reuse and discard) types under the same framework.
    
    \item \textbf{Full mechanisation in Agda}. The formalisation of the \picalc{} with leftover typing, from the syntax to the semantics and the type system, has been fully mechanised in Agda in about 1850 lines of code, and is publicly available at \cite{Zalakain2020Agda}.  We have fully mechanised all meta-theory and the details of a proof of subject reduction can be found in Appendix \ref{app:type-safety}.
\end{enumerate}

We use type level {de Bruijn indices} \cite{deBruijn1972,Dybjer1994} to define a syntax of \picalc{} processes that is \emph{well scoped by construction}: every free variable is accounted for in the type of the process that uses it (\autoref{syntax}).
We then provide an operational semantics for the \picalc{}, prior to any typing (\autoref{semantics}).
This operational semantics is defined as a reduction relation on processes.
The reduction relation tracks at the type level the channel on which communication occurs.
This information is later used to state the subject reduction theorem.
The reduction relation is defined modulo \emph{structural congruence} --- a relation defined on processes that acts as a quotient type to remove unnecessary syntactic minutiae introduced by the syntax of the \picalc{}.
We then define an interface for resource-aware algebras (\autoref{multiplicities}) and use it to parametrise a type system based on leftover typing (\autoref{leftover-typing}).
Finally, we present the meta theoretical properties of our type system in \autoref{meta-theory}.

\paragraph{Notation}

Data type definitions ($\type{\N}$) use double inference lines and index-free synonyms (\textsc{Nat}) as rule names for ease of reference.
Constructors ($\constr{0}$ and $\suc$) are used as inference rule names.
We maintain a close correspondence between the definitions presented in this paper and our mechanised definitions in Agda: inference rules become type constructors, premises become argument types and conclusions return types.
Universe levels and universe polymorphism are omitted for brevity --- all our types are of type $\Set$.
Implicit arguments are mentioned in type definitions but omitted by constructors.
\begin{mathpar}
  \datatype
  { }
  {\type{\N} : \Set}
  \; \textsc{Nat}

  \inferrule
  { }
  {\constr{0} : \type{\N}}

  \inferrule
  {n : \type{\N}}
  {\suc n : \type{\N}}
\end{mathpar}

We use colours to further distinguish the different entities in this paper.
$\type{TYPES}$ are blue and uppercased, with indices as subscripts, $\constr{constructors}$ are orange, $\func{functions}$ are teal, variables are black, and some constructor names are overloaded --- and disambiguated by context.

\section{Syntax}
\label{syntax}

In order to mechanise the \picalc{} syntax in Agda, we need to deal with bound names in continuation processes.
Names are cumbersome to mechanise: they are not inherently well scoped, one has to deal with alpha-conversion, and inserting new variables into a context entails proving that their names differ from all other names in context.
To overcome these challenges, we use de Bruijn indices \cite{deBruijn1972}, where a natural number $n$ (aka \emph{index}) is used to refer to the variable introduced $n$ binders ago.
That is, binders no longer introduce names; terms at different \emph{depths} use different indices to refer to the same binding.

While de Bruijn indices are useful for mechanisation, they are not as readable as names. To overcome this difficulty and demonstrate the correspondence between a \picalc{} that uses names and one that uses de Bruijn indices,
we provide \emph{conversion functions} in both directions and prove that they are inverses of each other up to $\alpha$-conversion. Further details can be found in Appendix~\ref{from_to_deBruijn}.

\begin{nidefinition}[\textsc{Var} and \textsc{Process}]\label{def:syntax}
A variable reference occurring under $n$ binders can refer to $n$ distinct variables.
We introduce the indexed family of types \cite{Dybjer1994} $\Var_n$: for all naturals $n$, the type $\Var_n$ has $n$ distinct elements.
We index processes according to their \emph{depth}: for all naturals $n$, a process of type $\Process_n$ contains free variables that can refer to $n$ distinct elements.
Every time we go under a binder, we increase the index of the continuation process, allowing the variable references within to refer to one more thing.
\begin{mathpar}
  \datatype
  {n : \N}
  {\Var_n : \Set}
  \; \textsc{Var}

  \inferrule
  {n : \N}
  {\constr{0} : \Var_{\suc n}}

  \inferrule
  {x : \Var_n}
  {\suc x : \Var_{\suc n}}

  \datatype
  {n : \N}
  {\Process_n : \Set}
  \; \textsc{Process}
\end{mathpar}

\begin{equation*}
  \begin{aligned}
    \Process_n ::=& \; \PO                &&\text{(inaction)}   \\ 
    |& \; \new{} \Process_{\suc n}         &&\text{(restriction)}          \\ 
    |& \; \comp{\Process_n}{\Process_n}    &&\text{(parallel)} \\ 
    |& \; \recv{\Var_n}{}\Process_{\suc n} &&\text{(input)}                \\ 
    |& \; \send{\Var_n}{\Var_n}\Process_n  &&\text{(output)}                  
  \end{aligned}
\end{equation*}

Process $\PO$ denotes the terminated process, where no further communications can occur;
process $\new{} P$ creates a new channel and binds it at index $0$ in the continuation process $P$;
process $\comp{P}{Q}$ composes $P$ and $Q$ in parallel;
process $\recv{x}{}P$ receives data along channel $x$ and makes that data available at index $0$ in the continuation process $P$;
process $\send{x}{y}{P}$ sends variable $y$ over channel $x$ and continues as process $P$.
\end{nidefinition}

\begin{example}[The courier system]
\label{example-process}

We present a courier system that consists of three \emph{roles}:
a sender, who wants to send a package;
a receiver, who receives the package sent by the sender;
and a courier, who carries the package from the sender to the receiver.
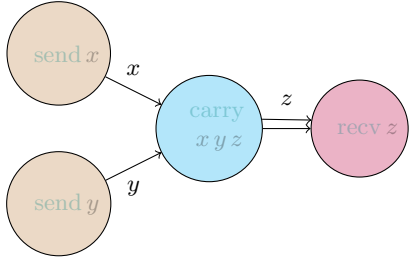
\begin{center}
  \small
  \begin{tikzpicture}[draw,circle]
    \node[draw,circle,fill opacity=0.3,text opacity=1,fill=brown] (send-x) at (0,2)   {$\func{send} \, x$};
    \node[draw,circle,fill opacity=0.3,text opacity=1,fill=brown] (send-y) at (0,0)   {$\func{send} \, y$};
    \node[draw,circle,fill opacity=0.3,text opacity=1,fill=purple] (recv-z) at (4,1)   {$\func{recv} \, z$};
    \node[draw,circle,fill opacity=0.3,text opacity=1,fill=cyan,align=center] (sync) at (2,1)   {$\func{carry}$\\$ \, x \, y \, z$};

    \draw[->] (send-x) -- node[above] {$x$} (sync);
    \draw[->] (send-y) -- node[below] {$y$} (sync);
    \draw[->] (sync.10) -- node[above] {$z$} (recv-z.170);
    \draw[->] (sync) -- (recv-z);
  \end{tikzpicture}
\end{center}

Our courier system is defined by four \picalc{} processes composed in parallel instantiating the above three roles:
we have two sender processes, $\func{send} \, x$ and $\func{send} \, y$, sending data over channels $x$ and $y$, respectively;
one receiver process, $\func{recv} \, z$, which receives over channel $z$ the data sent from each of the senders -- hence receives twice;
and a courier process $\func{carry} \, x \, y \, z$, which synchronises communication among the senders and the receiver.
The courier process first receives data from the two senders along its input channels $x$ and $y$, and then sends the two received bits of data to the receiver along its output channel $z$.

The sender and receiver \emph{roles} are defined below, parametrised by the channels on which they operate.
The sender creates a new channel to be sent as data, and sends it over channel $c$, and then terminates.
Processes $\func{send} \, x$ and $\func{send} \, y$ are an instantiation of $\func{send} \, c$.
The receiver receives data \emph{twice} on a channel $c$ and then terminates.
The receiver process $\func{recv} \, z$ is an instantiation of $\func{recv} \, c$.
\begin{align*}
  & \func{send} \; c = \new{} (\send{\suc c}{\constr{0}} \PO) &&
    \func{recv} \; c \; = \recv{c}{} \recv{(\suc c)}{} \PO
\end{align*}

The courier role is defined below as $\func{carry} \; x \; y \; z$.
It sequentially receives on the two input channels $x$ and $y$, instantiated as $in0$ and $in1$, and then outputs the two pieces of received data on the output channel $z$, instantiated as $out$.
Finally, we create three communication channels and compose all four processes together: the first channel is shared between the one sender and the courier, the second between the other sender and the courier, and the third between the receiver  and the courier.
The result is the courier $\func{system}$ defined below.
\begin{align*}
  & \func{carry} \; in0 \; in1 \; out \; =  \recv{in0 }{} 
  \recv{(\suc \; in1) }{} 
  \send{(\suc \suc \; out) }{\suc \constr{0}} 
  \send{(\suc \suc \; out) }{\constr{0}} \PO \\
  & \func{system} =  \new{} ( \func{send} \; \constr{0} 
  \comp{}{\new{} ( \func{send} \; \constr{0} } 
  \comp{}{\new{} ( \func{recv} \; \constr{0} } 
  \comp{}{\func{carry} \; (\suc \suc \constr{0}) \; (\suc \constr{0}) \; \constr{0}))) } 
\end{align*}

We continue this running example in  \autoref{leftover-typing}, where we provide typing derivations for the above processes and use a mix of linear, graded and shared typing to type the courier $\func{system}$.
\end{example}

\section{Operational Semantics}
\label{semantics}
Thanks to our well-scoped grammar in \autoref{syntax}, we now define the semantics of our language on the totality of the syntax.

\begin{nidefinition}[\textsc{Unused}] \label{def:unused}
  We consider a variable $i$ to be unused in $P$ ($\Unused_i \; P$) if none of the inputs nor the outputs refer to it.
  $\Unused_i \; P$ is defined as a recursive predicate on $P$, incrementing $i$ every time we go under a binder, and using $i \type{\not\equiv} x$ ( which unfolds to the negation of propositional equality on \textsc{Var}, i.e. $i \type{\equiv} x \to \type{\bot}$) to compare variables.
\end{nidefinition}

\begin{nidefinition}[\textsc{StructCong}] \label{def:structcong}
We define the base cases of a structural congruence relation $\eqeq$ as follows:
\begin{mathpar}
  \datatype
  { }
  {P \eqeq Q : \Set}
  \; \textsc{StructCong}

  \inferrule
  { }
  {\constr{comp-assoc} : \comp{P}{(\comp{Q}{R})} \eqeq \comp{(\comp{P}{Q})}{R}}

  \inferrule
  { }
  {\constr{comp-sym} : \comp{P}{Q} \eqeq \comp{Q}{P}}
  
  \inferrule
  { }
  {\constr{comp-id} : \comp{P}{\PO_n} \eqeq P}
  
  \inferrule
  { }
  {\constr{scope-end} : \new \PO_{\suc n} \eqeq \PO_n}
  
  \inferrule
  {uQ : \Unused_{\constr{0}} \; Q}
  {\constr{scope-ext} : \new (\comp{P}{Q}) \eqeq \comp{(\new P)}{\func{lower}_{\constr{0}} \; \; Q \; uQ}}

  \inferrule
  { }
  {\constr{scope-comm} : \new \new P \eqeq \new \new \func{exchange}_{\constr{0}} \; P}
\end{mathpar}
The first three rules ($\constr{comp}-$*) state associativity, symmetry, and $\PO$ as being the neutral element of parallel composition, respectively.
The last three ($\constr{scope}-$*) state garbage collection, scope extrusion and commutativity of restrictions, respectively.
In $\constr{scope-ext}$ the side condition $\Unused_i \; Q$ makes sure that $i$ is unused in $Q$ (see \autoref{def:unused}).
The function $\func{lower}_i \; Q \; uQ$ traverses $Q$ decrementing every index greater than $i$.
In $\constr{scope-comm}$ the function $\func{exchange}_i \; P$ traverses $P$ (of type $\allowbreak \Process_{\suc \suc n}$) and swaps variable references $i$ and $\suc i$.
In all the above, $i$ is incremented every time we go under a binder.
\end{nidefinition}

\begin{nidefinition}[\textsc{Equals}] \label{def:equals}
We lift the relation \textsc{StructCong} $\eqeq{}$ and close it under equivalence and congruence in $\eq{}$.
This relation is structurally congruent under a context $\mathcal{C}[\cdot]$ \cite{Sangio01} and is reflexive, symmetric and transitive. 
\end{nidefinition}

\begin{nidefinition}[\textsc{Reduces}]\label{def:reduces}
The operational semantics of the \picalc{} is defined as a reduction relation $\reduce{c}$ indexed by the channel $c$ on which communication occurs.
We keep track of channel $c$ so we can state subject reduction (\autoref{thm:subject-reduction}).

\begin{mathpar}
  \datatype
  {n : \N}
  {\Channel_n : \Set}
  \; \textsc{Channel}

  \inferrule
  { }
  {\constr{internal} : \Channel_n}

  \inferrule
  {i : \Var_n}
  {\constr{external} \; i : \Channel_n}

  \datatype
  {c : \Channel_n \\ P \; Q : \Process_n}
  {P \reduce{c} Q : \Set}
  \; \textsc{Reduces}

  \inferrule
  {i \; j : \Var_n \\ P : \Process_{\suc n} \\ Q : \Process_n}
  {\constr{comm} : \comp{\recv{i}{}P}{\send{i}{j}{Q}} \reduce{\constr{external} \; i} \comp{\func{lower}_{\constr{0}} \; (\subst{P}{\suc j}{\constr{0}}) \; uP'}{Q}}

  \inferrule
  {red : P \reduce{c} P'}
  {\constr{par} \; red : \comp{P}{Q} \reduce{c} \comp{P'}{Q}}

  \inferrule
  {red : P \reduce{c} Q}
  {\constr{res} \; red : \new P \reduce{\func{dec}\; c} \new Q}

  \inferrule
  {eq_1 : P \eq{} P' \\ red : P' \reduce{c} Q' \\ eq_2 : Q' \eq{} Q}
  {\constr{struct} \; eq \; red : P \reduce{c} Q}
\end{mathpar}

We distinguish between channels that are created inside the process ($\constr{internal}$), and channels that are created outside ($\constr{external}\ i$), where $i$ is the index of the channel variable.
In rule $\constr{comm}$, parallel processes reduce when they communicate over a common channel with index ${i}$.
As a result of that communication, the continuation of the input process $P$ has all the references to its most immediate variable substituted with references to $\suc j$, the variable sent by the output process $\send{i}{j}{Q}$.
After this substitution, $\subst{P}{\suc j}{\constr{0}}$ is \emph{lowered} --- all variable references are decreased by one (and we derive the proof $\Unused_{\constr{0}} \; (\subst{P}{\suc j}{\constr{0}})$).
Reduction is closed under parallel composition (rule $\constr{par}$), restriction (rule $\constr{res}$) and structural congruence (rule $\constr{struct}$) 
--- notably, not under input nor output, as doing so would not preserve the sequencing of actions \cite{Sangio01}.
Rule $\constr{res}$ uses $\func{dec}$ to decrement the index of channel $c$ as we wrap processes $P$ and $Q$ inside a binder.
It is defined as expected below:
\begin{equation*}
  \begin{aligned}
    & \func{dec} \; \constr{internal}                 && = \constr{internal} \\
    & \func{dec} \; (\constr{external} \; \constr{0}) && = \constr{internal} \\
    & \func{dec} \; (\constr{external} \; (\suc n))   && = \constr{external} \; n \\
  \end{aligned}
\end{equation*}
\end{nidefinition}

\section{Resource-aware Type System}
\label{type-system}

In \autoref{multiplicities} we characterise a usage algebra for our type system.
It defines how resources are \emph{split} in parallel composition and \emph{consumed} in input and output.
We define typing and usage contexts in \autoref{contexts}.
We provide a type system for a resource-aware \picalc{} in \autoref{leftover-typing}.

\subsection{Multiplicities and Capabilities}
\label{multiplicities}

In the linear \picalc{} each channel has an input and an output \emph{capability}, and each capability has a given \emph{multiplicity} of 0 (exhausted) or 1 (available).
We generalise over this notion by defining an algebra for multiplicities \cite{JungSSSTBD15,TuronTABD13} that is satisfied by linear, graded and shared types alike.
We then use pairs of multiplicities as usage annotations for a channel's input and output capabilities.

\begin{nidefinition}[\textsc{Algebra}]
  A \emph{usage algebra} is a ternary relation $\op{x}{y}{z}$ that is \emph{partial} (as not any two multiplicities can be combined), \emph{deterministic} and \emph{cancellative} (to aid equational reasoning) and \emph{associative} and \emph{commutative} (following directly from subject congruence for parallel composition).
  In addition, we ask that the leftovers can be \emph{computed} so that we can automatically update the usage context every time input and output occurs --- this is purely for usability.
  It has a \emph{neutral element} $\zero$ that is absorbed on either side, and that is also \emph{minimal} (so that new resources cannot arbitrarily spring into life).
  It has an element $\one$ that is used to count inputs and outputs.
  Below we define such an algebra as a record $\Algebra_C$ on a carrier $C$.
  (We use $\forall$ for universal quantification.
  The dependent product $\type{\exists}$ uses the value of its first argument in the type of its second.
  The type $\type{DEC} \; P$ is a witness of either $P$ or $P \to \type{\bot}$, where $\type{\bot}$ is the empty type with no constructors.)
  \begin{equation*}
  \begin{aligned}
    &\op{\_}{\_}{\_}         &:{} &                 &        & C \to C \to C \to \Set \\
    &\func{\cdot-unique}     &:{} &\forall x x' y z  & \to \; & \op{x'}{y}{z} \to \op{x}{y}{z} \to x' \equiv x \\
    &\func{\cdot-unique^l}   &:{} &\forall x y y' z  & \to \; & \op{x}{y'}{z} \to \op{x}{y}{z} \to y' \equiv y \\
    &\func{\cdot-assoc}      &:{} &\forall x y z u v & \to \; & \op{x}{y}{z} \to \op{y}{u}{v} \to \type{\exists} w  \; (\op{x}{u}{w} \times \op{w}{v}{z}) \\
    &\func{\cdot-comm}       &:{} &\forall x y z     & \to \; & \op{x}{y}{z} \to \op{x}{z}{y} \\
    &\func{\cdot-compute^r}  &:{} &\forall x y       & \to \; & \type{DEC} \; (\type{\exists} z  \; (\op{x}{y}{z})) \\
    &\zero                   &:{} &                 &        & C \\
    &\func{\cdot-id^l}       &:{} &\forall x         & \to \; & \op{x}{\zero}{x} \\
    &\func{\cdot-min^l}      &:{} &\forall y z        & \to \; & \op{\zero}{y}{z} \to y \equiv \zero \\
    &\one                    &:{} &                 &        & C \\
  \end{aligned}
  \end{equation*}
\end{nidefinition}

We sketch the implementation of linear, graded and shared types as instances of our usage algebra below.
Their use in typing derivations is illustrated in \autoref{example-derivation}.
\begin{center}
\begin{tabular}{l | l | l}
  & carrier & operation \\
  \hline
  \textbf{linear} & \makecell[cl]{$\constr{0} \, : \, \type{Lin}$ \\ $\constr{1} \, : \, \type{Lin}$} & \makecell[cl]{$\op{\constr{0}}{\constr{0}}{\constr{0}}$ \\ $\op{\constr{1}}{\constr{1}}{\constr{0}}$ \\ $\op{\constr{1}}{\constr{0}}{\constr{1}}$} \\
  \hline
  \textbf{graded} & \makecell[cl]{$\constr{0} \, : \type{Gra}$ \\ $\suc \, : \type{Gra} \to \type{Gra}$} & \makecell[cl]{$\forall \, x \, y \, z$ \\ $\to x \, \type{\equiv} \, y \, \func{+} \, z$ \\ $\to \op{x}{y}{z}$} \\
  \hline
  \textbf{shared} & $\constr{\omega} \, : \, \type{Sha}$ & $\op{\constr{\omega}}{\constr{\omega}}{\constr{\omega}}$ \\
\end{tabular}
\end{center}

\subsection{Typing Contexts}
\label{contexts}

We use indexed sets of usage algebras to allow several usage algebras to coexist in our type system with leftovers (\autoref{leftover-typing}).
\begin{nidefinition}[\textsc{Algebras}]
  An \emph{indexed set of usage algebras} is a type $\Idx$ of indices that is nonempty ($\type{\exists IDX}$) together with an interpretation $\Usage$ of indices into types, and an interpretation $\type{ALGEBRAS}$ of indices into usage algebras of the corresponding type.

  \begin{equation*}
    \begin{aligned}
      &\Idx               &:{} &\Set \\
      &\type{\exists IDX} &:{} &\Idx \\
      &\Usage             &:{} &\Idx \to \Set \\
      &\type{ALGEBRAS}    &:{} &(idx : \Idx) \to \Algebra_{\Usage_{idx}}
    \end{aligned}
  \end{equation*}
\end{nidefinition}

We keep typing contexts ($\PreCtx$) and usage contexts ($\Ctx$) separate.
The former are preserved throughout typing derivations; the latter are transformed as a result of input, output, and context splits.
\begin{nidefinition}[\textsc{Type} and \textsc{PreCtx}: types and typing contexts]
  A \emph{type} is either a unit type ($\unit$),
  or a channel type ($\channel{t}{x}$).
  \begin{mathpar}
    \datatype
    { }
    {\Type : \Set}
    \; \textsc{Type}

    \inferrule
    { }
    {\unit : \Type}

    \inferrule
    {t : \Type \\ \stacked{idx : \Idx \\\\ x : \Usage_{idx}^{\func{2}}}}
    {\channel{t}{x} : \Type}
  \end{mathpar}

  The unit type $\unit$ serves as a base case for types.
  The type $\channel{t}{x}$ of a channel determines what type ${t}$ of data and what usage annotations ${x}$ are sent over that channel --- we use the notation $\type{C}^{\func{2}}$ to stand for a $\type{C} \constr{\times} \type{C}$ pair of input and output multiplicities, respectively.
  This channel notation aligns with $[t]\ \mathbf{chan}_{(i^{y},o^{z})}$, where $y,z$ are the input and output multiplicities, respectively \cite{K07}.
  Henceforth, we use $\lz$ to denote the multiplicity pair $\zero \comma \zero$, $\li$ for the pair $\one \comma \zero$, $\lo$ for $\zero \comma \one$, and $\lio$ for $\one \comma \one$.
  This notation was originally used in the linear \picalc{} \cite{KPT96,Sangio01}.
  A \emph{typing context} $\PreCtx_n$ is a length-indexed list of types that is either empty ($\constr{[]}$) or the result of appending a type $t : \Type$ to an existing context ($\gamma \constr{,} t$).
\end{nidefinition}

\begin{nidefinition}[\textsc{Idxs} and \textsc{Ctx}: contexts of indices and usage contexts]
  A context of indices $\Idxs_n$ is a length-indexed list that is either empty ($\constr{[]}$) or the result of appending an index $i : \Idx$ to an existing context ($idxs \constr{,} i$).
  A \emph{usage context} is a context $\Ctx_{idxs}$ indexed by a context of indices $idxs : \Idxs_n$ that is either empty ($\constr{[]}$) or the result or appending a usage annotation pair $u : \Usage_{idx}^{\func{2}}$ with index $idx : \Idx$ to an existing context ($\Gamma \constr{,} u$).
\end{nidefinition}

\subsection{Typing with Leftovers}
\label{leftover-typing}

We present a resource-aware type system for the \picalc{} based on \emph{leftover typing} \cite{Allais2018a}, a technique that, in addition to the usual typing context $\PreCtx_n$ and (input) usage context $\Ctx_{idxs}$, adds an extra \emph{(output) usage context} $\Ctx_{idxs}$ to the typing rules.
This output context contains the \emph{leftovers} (the unused multiplicities) of the process being typed.
These leftovers can then be used as input to another typing derivation.

Leftover typing inverts the information flow of usage annotations so that it is the typing derivations of subprocesses which determine how resources are allocated.
As a result, context split proofs are no longer necessary.
Leftover typing also allows \emph{framing} to be stated, and \emph{weakening} and \emph{strengthening} to cover linear types too.

Our type system is composed of two typing judgments: one for variable references (\autoref{def:varref}) and one for processes (\autoref{def:types}).
Both judgments are indexed by a typing context $\gamma$, an input usage context $\Gamma$, and an output usage context $\Delta$ (the leftovers).
The \textbf{typing judgement for variables} $\contains{\gamma}{\Gamma}{i}{t}{y}{\Delta}$ asserts that ``index $i$ in typing context $\gamma$ is of type $t$, and subtracting $y$ at position $i$ from input usage context $\Gamma$ results in leftovers $\Delta$''.
The \textbf{typing judgement for processes} $\types{\gamma}{\Gamma}{P}{\Delta}$ asserts that ``process $P$ is well typed under typing context $\gamma$, usage input context $\Gamma$ and leftovers $\Delta$''.

\begin{nidefinition}[\textsc{VarRef}: typing variable references] \label{def:varref}
  The \textsc{VarRef} typing relation for variable references is presented below.
  \begin{mathpar}
    \mprset{sep=0.5em}
  
    \datatype{
      \gamma : \PreCtx_n \\
      i : \Var_n \\
      \stacked{
        t : \Type \\\\
        idx : \Idx \\\\
        y : \Usage_{idx}^{\func{2}}} \\
      \stacked{
        idxs : \Idxs_n \\\\
        \Gamma \; \Delta : \Ctx_{idxs}}}
    {\contains{\gamma}{\Gamma}{i}{t}{y}{\Delta} : \Set}
    \; \textsc{VarRef}
  
    \inferrule
    {\opsquared{x}{y}{z}}
    {\constr{0} : \contains{\gamma \comma t}{\Gamma \comma x}{\constr{0}}{t}{y}{\Gamma \comma z}}
    
    \inferrule
    {v : \contains{\gamma}{\Gamma}{i}{t}{x}{\Delta}}
    {\suc \; v : \contains{\gamma \comma t'}{\Gamma \comma x'}{\suc i}{t}{x}{\Delta \comma x'}}
  \end{mathpar}

  We lift the operation $\op{x}{y}{z}$ and its algebraic properties to an operation $\opsquared{(x_l \comma x_r)}{(y_l \comma y_r)}{(z_l \comma z_r)}$ on pairs of multiplicities.
  The base case $\constr{0}$ splits the usage annotation $x$ of type $\Usage_{idx}^{\func{2}}$ into $y$ and $z$ (the leftovers).
  Note that the remaining context $\Gamma$ is preserved unused as a leftover.
  This splitting $\opsquared{x}{y}{z}$ is as per the usage algebra provided by the developer for the index $idx$.
  In our Agda implementation, $\opsquared{x}{y}{z}$ is actually a trivially satisfiable implicit argument if $\opsquared{x}{y}{z}$ is inhabited and an unsatisfiable argument otherwise.
  The inductive case $\suc$ appends the type $t'$ to the typing context, and the usage annotation $x'$ to both the input and output usage contexts. 

\end{nidefinition}

\begin{example}[Variable reference]
  $\func{egVar}$ defines a variable reference $\suc \constr{0}$ with type $\channel{\unit}{\li}$ and usage $\li$.
  We must show that this variable is well typed in an environment with a typing context $\gamma = \constr{[]} \comma \channel{\unit}{\li} \comma \unit$ and a usage context $\Gamma = \constr{[]} \comma \lio \comma \lio$.
  The \textsc{VarRef} constructors are completely determined by the variable index $\suc \; \constr{0}$ in the type.
  The constructor $\suc$ steps under the outermost variable in the context, preserving its usage annotation $\lio$ from input to output.
  The constructor $\constr{0}$ asserts that the next variable is of type $\channel{\unit}{\li}$, and that the usage annotation $\lio$ can be split such that $\op{\lio}{\li}{\lo}$ --- using $\func{\cdot-compute^r}$ to automatically fulfill the proof obligation.
  \begin{flalign*}
  & \func{egVar} \; : \; \contains{(\constr{[]} \comma \channel{\unit}{\li} \comma \unit)} {(\constr{[]} \comma \lio \comma \lio)} {\suc \constr{0}} {\channel{\unit}{\li}} {\li} {(\constr{[]} \comma \lo \comma \lio)}\\
  & \func{egVar} \; = \; \suc \constr{0}
  \end{flalign*}
\end{example}

\begin{nidefinition}[\textsc{Types}: typing processes] \label{def:types}
  The \textsc{Types} typing relation for the resource-aware \picalc{} processes is presented below.
  For convenience, we reuse the constructor names introduced for the syntax in \autoref{syntax}.
  \begin{mathpar}
    \datatype{
      \gamma : \PreCtx_n \\
      P : \Process_n \\
      \stacked{
        idxs : \Idxs_n \\\\
        \Gamma \; \Delta : \Ctx_{idxs}}}
    {\types{\gamma}{\Gamma}{P}{\Delta} : \Set}
    \; \textsc{Types}
    
    \inferrule
    { }
    {\PO : \types{\gamma}{\Gamma}{\PO}{\Gamma}}
  
    \inferrule
    {t : \Type \\ x : \Usage_{idx}^{\func{2}} \\ y : \Usage_{idx'} \\\\
     cont : \types{\gamma \comma \channel{t}{x}}{\Gamma \comma (y \comma y) }{P}{\Delta \comma \lz}}
    {\new \; t \; x \; y \; cont : \types{\gamma}{\Gamma}{\new P}{\Delta}}
  
    \inferrule
        {\stacked{
            chan : \contains{\gamma \hspace{1.0em}}{\Gamma \hspace{1.0em}}{i}{\channel{t}{x}}{\li}{\Xi} \\\\
            cont \hspace{0.3em} : \types{\gamma \comma t}{\Xi \comma x}{P \hspace{3.8em}}{\Theta \comma \lz}}}
        {\recv{chan}{} cont : \types{\gamma}{\Gamma}{\recv{i}{}P}{\Theta}}
  
    \inferrule
        {\stacked{
            chan : \contains{\gamma}{\Gamma \hspace{0.1em}}{i}{\channel{t}{x}}{\lo}{\Delta} \\\\
            loc \hspace{0.9em} : \contains{\gamma}{\Delta}{j}{t \hspace{2.8em}}{x \hspace{0.2em}}{\Xi} \\\\
            cont \hspace{0.3em} : \types{\gamma}{\Xi}{\hspace{0.4em}P\hspace{3.8em}}{\Theta}}}
        {\send{chan}{loc} cont : \types{\gamma}{\Gamma}{\send{i}{j}P}{\Theta}}
  
    \inferrule
    {l : \types{\gamma}{\Gamma}{P\hspace{0.3em}}{\Delta} \\\\
     r : \types{\gamma}{\Delta}{Q}{\Xi}}
    {\comp{l}{r} : \types{\gamma}{\Gamma}{\comp{P}{Q}}{\Xi}}
  \end{mathpar}

  The inaction process in rule $\PO$ does not change usage annotations.
  The scope restriction in rule $\new$ expects three arguments: the type $t$ of data being transmitted; the usage annotation $x$ of what is being transmitted; and the multiplicity $y$ given to the channel itself.
  This multiplicity $y$ is used for both input and output, so that they are balanced.
  The continuation process $P$ is provided with the new channel with usage annotation $y \comma y$, which it must completely exhaust.
  The input process in rule $\recv{}{}$ requires a channel $chan$ at index $i$ with usage $\li$ available, such that data with type $t$ and usage $x$ can be sent over it.
  Note that the index $i$ is determined by the syntax of the typed process.
  We use the leftovers $\Xi$ to type the continuation process, which is also provided with the received element --- of type $t$ and multiplicity $x$ --- at index $\constr{0}$.
  The received element $x$ must be completely exhausted by the continuation process.
  Similarly to input, the output process in rule $\send{}{}$ requires a channel $chan$ at index $i$ with usage $\lo$ available, such that data with type $t$ and usage $x$ can be sent over it.
  We use the leftover context $\Delta$ to type the transmitted data, which needs an element $loc$ at index $j$ with type $t$ and usage $x$, as per the type of the channel $chan$.
  The leftovers $\Xi$ are used to type the continuation process.
  Note that both indices $i$ and $j$ are determined by the syntax of the typed process.
  Parallel composition in rule $\comp{}{}$ uses the leftovers of the left-hand process to type the right-hand process.
  Indeed, \autoref{thm:subject-congruence} shows that an alternative rule where the resources are first threaded through $Q$ is admissible too.
\end{nidefinition}

\begin{example}[Typing derivation (Continued)]
\label{example-derivation}
We provide the typing derivation for the courier system defined in \autoref{example-process}.
For the sake of simplicity, we instantiate these processes with concrete variable references before typing them.

The receiver defined by the $\func{recv}$ process receives data along the channel with index $\constr{0}$, which needs to be of type $\channel{t}{u}$ for some $t$ and $u$.
After receiving twice, the process ends: we must not be left with any unused multiplicities, thus $u = \lz$.
We will use graded types to keep track of the exact number of times communication happens.
Whatever the input multiplicity of the channel, we will consume $2$ of it and leave the remaining as leftovers.
The sender defined by the $\func{send}$ process sends data along the channel with index $\constr{0}$, which needs to be of type $\channel{t}{u}$ for some $t$ and $u$.
We instantiate $t$ (the type of data that the sender sends) to the trivial channel $\channel{\unit}{\func{\omega}}$.
As per the type of the process $\func{recv}$, $u = \lz$.
We will transmit once, thus use $\suc{\constr{0}}$ output multiplicity, and leave the rest as leftovers.
Agda can uniquely determine the arguments required by the $\new{}$ constructor.
\begin{align*}
   & \func{recvwt} \;  : \; \types{\gamma \comma \channel{t}{\lz}  }{\Gamma \comma (\suc \suc l \comma r)  }{\func{recv} \; \constr{0}  }{\Gamma \comma (l \comma r)} \\
   & \func{recvwt} \;  = \; \recv{\constr{0}}{} \recv{(\suc \constr{0})}{} \PO \\
   & \func{sendwt} \;  : \; \types{\gamma \comma \channel{\channel{\unit}{\func{\omega}}}{\lz}  }{\Gamma \comma (l \comma \suc r)  }{\func{send} \; \constr{0}  }{\Gamma \comma (l \comma r)}  \\
   & \func{sendwt} \;  = \; \new{} \; \_ \; \_ \; \zero \; (\send{\suc \constr{0}}{\constr{0}} \PO)
\end{align*}

Dually, the courier defined by the $\func{carry}$ process expects input multiplicities for the channels shared with $\func{send}$ and output multiplicities for the channel shared with $\func{recv}$.
We can now compose these processes in parallel and type the courier $\func{system}$.
\begin{align*}
& \func{carrywt} && : \types{\gamma \comma \channel{t}{\lz} \comma \channel{t}{\lz} \comma \channel{t}{\lz}\\ & &&}{\Gamma \comma (\suc lx \comma rx) \comma (\suc ly \comma ry) \comma (lz \comma \suc \suc rz)\\ & &&}{\func{carry} \; (\suc \suc \constr{0}) \; (\suc \constr{0}) \; \constr{0}\\ & &&}{\Gamma \comma (lx \comma rx) \comma (ly \comma ry) \comma (lz \comma rz)} & \\
& \func{carrywt} && = \recv{(\suc \suc \constr{0})}{} \recv{(\suc \suc \constr{0})}{} \send{(\suc \suc \constr{0})}{\suc \constr{0}} \send{(\suc \suc \constr{0})}{\constr{0}} \PO \\
& \func{systemwt} && : \; \types{\constr{[]}}{\constr{[]}}{\func{system}}{\constr{[]}} \\
& \func{systemwt} && = \new{} \; \_ \; \_ \; \_ \; ( \func{sendwt} \comp{}{ \new{} \; \_ \; \_ \; \_ } \; ( \func{sendwt} \comp{}{ \new{} \; \_ \; \_ \; \_ } \; ( \func{recvwt} \comp{}{ \func{carrywt} })))
\end{align*}
\end{example}

\section{Meta-Theory} \label{meta-theory}

We have mechanised subject reduction for our \picalc{} with leftovers in 850 lines of Agda code.
The meta-theory of resource-aware type systems often needs to reason on typing derivations modulo associativity in the allocation of resources.
For type systems using context splitting side conditions, this means applying associativity lemmas to recompute context splits; for type systems using leftover typing it does not.
As an example, the proof that $\constr{comp-asssoc}$ preserves typing proceeds by deconstructing the input derivation into $\comp{P}{(\comp{Q}{R})}$ and reassembling it as $\comp{(\comp{P}{Q})}{R}$ without the need of any extra reasoning.

All the reasoning carried out in our type safety proofs is based on the algebraic properties introduced in \autoref{multiplicities} -- the exception to this is $\func{\cdot -compute^r}$, only there for the user's convenience.
  We lift the operation $\opsquared{x}{y}{z}$ and its algebraic properties to an operation $\opctx{\Gamma}{\Delta}{\Xi}$ on usage contexts that have the same underlying context of indices.
The algebraic properties of the algebras allow us to see a typing derivation $\types{\gamma}{\Gamma}{P}{\Delta}$ as a unique \emph{arrow} from $\Gamma$ to $\Delta$, and to freely compose and reason with arrows with the same typing context and a matching output and input usage contexts.

Leftover typing also allows us to state a \emph{framing} theorem showing that adding or subtracting arbitrary usage annotations to the input and output usage contexts preserves typing -- one can understand a typing derivation independently from its unused resources.
With framing one can show that $\constr{comp-comm}$ preserves typing: in $\comp{P}{Q}$ the typing of $P$ and $Q$ is independent of one another.
\begin{nitheorem}[Framing]
  \label{thm:framing}
  Let $\types{\gamma}{\Gamma_l}{P}{\Xi_l}$.
  Let $\Delta$ be such that $\opctx{\Gamma_l}{\Delta}{\Xi_l}$.
  Then for any $\Gamma_r$ and $\Xi_r$ where $\opctx{\Gamma_r}{\Delta}{\Xi_r}$ it holds that $\types{\gamma}{\Gamma_r}{P}{\Xi_r}$.
\end{nitheorem}

Leftover typing allows \emph{weakening} and \emph{strengthening} to acquire a more general form where linear variables can freely be added or removed from context too -- as long as they are added and removed to and from both the input and output contexts.

\begin{nitheorem}[Weakening]
  \label{thm:weakening}
  Let $\func{ins}_i$ insert an element into a context at position $i$.
  Let $P$ be well typed in $\types{\gamma}{\Gamma}{P}{\Xi}$.
  Then, lifting every variable greater than or equal to $i$ in $P$ is well typed in
  $\types{\func{ins}_i \; t \; \gamma}{\func{ins}_i \; x \; \Gamma}{\func{lift}_i \; P}{\func{ins}_i \; x \; \Xi}$.
\end{nitheorem}

\begin{nitheorem}[Strengthening]
  \label{thm:strengthening} 
  Let $\func{del}_i$ delete the element at position $i$ from a context.
  Let $P$ be well typed in $\types{\gamma}{\Gamma}{P}{\Xi}$.
  Let $i$ be a variable not in $P$, such that $uP \; : \; \Unused_i \; P$.
  Then lowering every variable greater than $i$ in $P$ is well typed in $\types{\func{del}_i \; \gamma}{\func{del}_i \; \Gamma}{\func{lower}_i \; P \; uP}{\func{del}_i \; \Xi}$.
\end{nitheorem}

Subject congruence states that structural congruence (\autoref{def:equals}) preserves the well-typedness of a process.
\begin{nitheorem}[Subject Congruence]
  \label{thm:subject-congruence}
  Let $P$ and $Q$ be processes. If $P \eq{} Q$ and $\types{\gamma}{\Gamma}{P}{\Xi}$, then $\types{\gamma}{\Gamma}{Q}{\Xi}$.
\end{nitheorem}

Finally, subject reduction states that reducing on a channel $c$ (\autoref{def:reduces}) preserves the well-typedness of a process --- after consuming $\lio$ from $c$ if $c$ is an $\constr{external}$ channel.  Below we use $\containsusage{\Gamma}{i}{x}{\Delta}$ to stand for $\contains{\gamma}{\Gamma}{i}{t}{x}{\Delta}$ for some $\gamma$ and $t$.
\begin{nitheorem}[Subject Reduction]
  \label{thm:subject-reduction}
  Let $\types{\gamma}{\Gamma}{P}{\Xi}$ and $P \reduce{c} Q$.
  If $c$ is $\constr{internal}$, then $\types{\gamma}{\Gamma}{Q}{\Xi}$.
  If $c$ is $\constr{external} \; i$ and $\containsusage{\Gamma}{i}{\lio}{\Delta}$, then $\types{\gamma}{\Delta}{Q}{\Xi}$.
\end{nitheorem}
We refer to Appendix \ref{app:type-safety} for a more detailed account of the mechanised proofs.

\section{Conclusions, Related and Future Work}

\paragraph*{Extrinsic Encodings}

Extrinsic encodings define a syntax (often well-scoped) and a runtime semantics prior to any type system.
This allows one to talk about ill-typed terms, and defers the proof of subject reduction to a later stage.
To the best of our knowledge, leftover typing makes its appearance in 1994, when Ian Mackie first uses it to formulate intuitionistic linear logic \cite{Mackie}.
Allais \cite{Allais2018a} uses leftover typing to mechanise in Agda a bidirectional type system for the linear \lambdacalc{}.
He proves type preservation and provides a decision procedure for type checking and type inference.
In this paper, we follow Allais \cite{Allais2018a} and apply leftover typing to the \picalc{} for the first time.
We generalise the usage algebra, leading to linear, graded and shared type systems.
Drawing from quantitative type theory (by McBride and Atkey \cite{McBride2016,Atkey2018}), in our work we too are able to talk about fully consumed resources --- e.g., we can transmit $\lz$ multiplicities of a fully exhausted channel.
Recent years have seen an increase in the efforts to mechanise resource-aware process algebras, but one of the earliest works is the mechanisation of the linear \picalc{} in Isabelle/HOL by Gay \cite{Gay2001}.
Gay encodes the \picalc{} with linear and shared types using de Bruijn indices, a reduction relation and a type system posterior to the syntax.
However, in his work typing rules demand user-provided context splits, and variables with consumed usage annotations are erased from context.
We remove the demand for context splits, preserve the ability to talk about consumed resources, and adopt a more general usage algebra.
Orchard et al. introduce Granule \cite{Orchard}, a fully-fledged functional language with graded modal types, linear types, indexed types and polymorphism.
Modalities include exact usages, security levels and intervals; resource algebras are pre-ordered semirings with partial addition.
The authors provide bidirectional typing rules, and show the type safety of their semantics.
The work by Goto et al. \cite{Goto2016a} is, to the best of our knowledge, the first formalisation of session types which comes along with a mechanised proof of type safety in Coq.
The authors extend session types with polymorphism and pattern matching.
They use a locally-nameless encoding for variable references, a syntax prior to types, and an LTS semantics that encodes session-typed processes into the \picalc{}.
Their type system uses reordering of contexts and extrinsic context splits, which are not needed in our work. 

\paragraph*{Intrinsic Encodings}
 
Intrinsic encodings merge syntax and type system.
As a result, one can only ever talk about well-typed terms, and the reduction relation by construction carries a proof of subject reduction.
Significantly, by merging the syntax and static semantics of the object language one can fully use the expressive power of the host language.
Thiemann formalises in Agda the MicroSession (minimal GV \cite{Gay2010}) calculus with support for recursion and subtyping \cite{Thiemann2019}.
As Gay does in \cite{Gay2001}, context splits are given extrinsically, and exhausted resources are removed from typing contexts altogether.
The runtime semantics are given as an intrinsically typed CEK machine with a global context of session-typed channels.
In their recent paper, Ciccone and Padovani mechanise a dependently-typed linear \picalc{} in Agda \cite{Ciccone}.
Their intrinsic encoding allows them to leverage Agda's dependent types to provide a dependently-typed interpretation of messages --- to avoid linearity violations the interpretation of channel types is erased.
Message input is modeled as a dependent function in Agda, and as a result message predicates, branching, and variable-length conversations can be encoded.
In contrast to our work, their algebra is on the multiplicities $0$, $1$, $\omega$, and top-down context splitting proofs must be provided.
In another recent work, Rouvoet et al. provide an intrinsic type system for a \lambdacalc{} with session types \cite{Rouvoet2020}.
They use proof relevant separation logic and a notion of a supply and demand \emph{market} to make context splits transparent to the user.
Their separation logic is based on a partial commutative monoid that need not be deterministic nor cancellative.
Their typing rules preserve the balance between supply and demand, and are extremely elegant.
They distill their typing rules even further by modelling the supply and demand market as a state monad.

\paragraph*{Other Work}

Castro et al. \cite{Castro2020} provide tooling for locally-nameless representations of process calculi in Coq, where de Bruijn indices are less popular than in Agda or Idris.
They use their tool to help automate proofs of subject reduction for a type system with session types.
Orchard and Yoshida \cite{OrchardY16} embed a small effecftul imperative language into the session-typed \picalc{}, showing that session types are expressive enough to encode effect systems.
Based on contextual type theory, LINCX \cite{Georges2017} extends the linear logical framework LLF \cite{Cervesato1996} by internalising the notion of bindings and contexts.
The result is a meta-theory in which HOAS encodings with both linear and dependent types can be described.
The developer obtains for free an equational theory of substitution and decidable typechecking without having to encode context splits within the object language.
Further work on mechanisation of the \picalc{} \cite{Henry-Gerard1999,Honsell2001a,Bengtson2013,Despeyroux2000,Affeldt2008}, focuses on non-linear variations, differently from our range of linear, graded and shared types.

\paragraph*{Conclusions and Future Work}

We provide a well-scoped syntax and a semantics for the \picalc{}, extrinsically define a type system on top of the syntax capable of handling linear, graded and shared types under the same unified framework and show subject reduction.
We avoid extrinsic context splits by defining a type system based on leftover typing \cite{Allais2018a}.
As a result, theorems like framing, weakening and strengthening can now be stated also for the linear \picalc{}.
Our work is fully mechanised in around 1850 lines of code in Agda \cite{Zalakain2020Agda}.

As future work we intend to expand our framework to include infinite behaviour by adding process replication, which is challenging, as to prove subject congruence one needs to uniquely determine the resources consumed by a process --- e.g., by adding type annotations to the syntax.
Orthogonally, we aim to investigate making our typing rules bidirectional which would allow us to provide a decision procedure for type checking processes in a given set of algebras.
Finally, we will use our \picalc{} with leftovers as an underlying framework on top of which we can implement session types, via their encodings into linear types \cite{Dardha14,DardhaGS17,ScalasDHY17} and other advanced type theories.

\paragraph*{Acknowledgments}

We want to thank Erika, Wen Kokke, James Wood, Guillaume Allais, Bob Atkey, and Conor McBride for their valuable suggestions.

\clearpage
\bibliographystyle{splncs04}
\bibliography{paper}

\begin{thebibliography}{10}
\providecommand{\url}[1]{\texttt{#1}}
\providecommand{\urlprefix}{URL }
\providecommand{\doi}[1]{https://doi.org/#1}

\bibitem{Affeldt2008}
Affeldt, R., Kobayashi, N.: {A Coq Library for Verification of Concurrent
  Programs}. Electron. Notes Theor. Comput. Sci.  \textbf{199},  17--32 (2008).
  \doi{10.1016/j.entcs.2007.11.010}

\bibitem{Allais2018a}
Allais, G.: {Typing with Leftovers - A mechanization of Intuitionistic
  Multiplicative-Additive Linear Logic}. In: Types for Proofs and Programs,
  {TYPES}. LIPIcs, vol.~104, pp. 1:1--1:22. Schloss Dagstuhl - Leibniz-Zentrum
  f{\"{u}}r Informatik (2017). \doi{10.4230/LIPIcs.TYPES.2017.1}

\bibitem{Atkey2018}
Atkey, R.: {Syntax and Semantics of Quantitative Type Theory}. In: Logic in
  Computer Science, {LICS}. pp. 56--65. {ACM} (2018).
  \doi{10.1145/3209108.3209189}

\bibitem{BarendsenS96}
Barendsen, E., Smetsers, S.: {Uniqueness Typing for Functional Languages with
  Graph Rewriting Semantics}. Math. Struct. Comput. Sci.  \textbf{6}(6),
  579--612 (1996)

\bibitem{Bengtson2013}
Bengtson, J.: {The pi-calculus in nominal logic}, vol.~2012 (2012),
  \url{https://www.isa-afp.org/entries/Pi\_Calculus.shtml}

\bibitem{Castro2020}
Castro, D., Ferreira, F., Yoshida, N.: {EMTST: Engineering the Meta-theory of
  Session Types}. In: Tools and Algorithms for the Construction and Analysis of
  Systems, {TACAS}. Lecture Notes in Computer Science, vol. 12079, pp.
  278--285. Springer (2020). \doi{10.1007/978-3-030-45237-7\_17}

\bibitem{Cervesato1996}
Cervesato, I., Pfenning, F.: {A Linear Logical Framework}. In: Logic in
  Computer Science, {LICS}. pp. 264--275. {IEEE} Computer Society (1996).
  \doi{10.1109/LICS.1996.561339}

\bibitem{Ciccone}
Ciccone, L., Padovani, L.: {A Dependently Typed Linear {\(\pi\)}-Calculus in
  Agda}. In: {PPDP} '20: 22nd International Symposium on Principles and
  Practice of Declarative Programming. pp. 8:1--8:14. {ACM} (2020).
  \doi{10.1145/3414080.3414109}

\bibitem{Dardha14}
Dardha, O.: {Recursive Session Types Revisited}. In: Carbone, M. (ed.) Workshop
  on Behavioural Types, {BEAT}. {EPTCS}, vol.~162, pp. 27--34 (2014).
  \doi{10.4204/EPTCS.162.4}

\bibitem{DardhaGS12}
Dardha, O., Giachino, E., Sangiorgi, D.: {Session types revisited}. In:
  Principles and Practice of Declarative Programming, {PPDP}. pp. 139--150.
  {ACM} (2012). \doi{10.1145/2370776.2370794}

\bibitem{DardhaGS17}
Dardha, O., Giachino, E., Sangiorgi, D.: {Session types revisited}. Inf.
  Comput.  \textbf{256},  253--286 (2017). \doi{10.1016/j.ic.2017.06.002},
  extended version of \cite{DardhaGS12}

\bibitem{deBruijn1972}
{de Bruijn}, N.G.: {Lambda Calculus Notation with Nameless Dummies, a Tool for
  Automatic Formula Manipulation, with Application to the Church-Rosser
  Theorem}. In: Indagationes {{Mathematicae}} ({{Proceedings}}). vol.~75, pp.
  381--392. {Elsevier} (1972)

\bibitem{Henry-Gerard1999}
Deransart, P., Smaus, J.: {Subject Reduction of Logic Programs as
  Proof-Theoretic Property}, vol.~2002 (2002),
  \url{http://danae.uni-muenster.de/lehre/kuchen/JFLP/articles/2002/S02-01/JFLP-A02-02.pdf}

\bibitem{Despeyroux2000}
Despeyroux, J.: {A Higher-Order Specification of the pi-Calculus}, Lecture
  Notes in Computer Science, vol.~1872. Springer (2000).
  \doi{10.1007/3-540-44929-9\_30}

\bibitem{Dybjer1994}
Dybjer, P.: {Inductive Families}. Formal Asp. Comput.  \textbf{6}(4),  440--465
  (1994). \doi{10.1007/BF01211308}

\bibitem{Gay2001}
Gay, S.J.: {A Framework for the Formalisation of Pi Calculus Type Systems in
  Isabelle/HOL}. In: Theorem Proving in Higher Order Logics, {TPHOLs}. Lecture
  Notes in Computer Science, vol.~2152, pp. 217--232. Springer (2001).
  \doi{10.1007/3-540-44755-5\_16}

\bibitem{Gay2010}
Gay, S.J., Vasconcelos, V.T.: {Linear type theory for asynchronous session
  types}. J. Funct. Program.  \textbf{20}(1),  19--50 (2010).
  \doi{10.1017/S0956796809990268}

\bibitem{Georges2017}
Georges, A.L., Murawska, A., Otis, S., Pientka, B.: {LINCX: A Linear Logical
  Framework with First-Class Contexts}. In: European Symposium on Programming,
  {ESOP}, Lecture Notes in Computer Science, vol. 10201, pp. 530--555. Springer
  (2017). \doi{10.1007/978-3-662-54434-1\_20}

\bibitem{Goto2016a}
Goto, M.A., Jagadeesan, R., Jeffrey, A., Pitcher, C., Riely, J.: {An extensible
  approach to session polymorphism}. Math. Struct. Comput. Sci.
  \textbf{26}(3),  465--509 (2016). \doi{10.1017/S0960129514000231}

\bibitem{Honsell2001a}
Honsell, F., Miculan, M., Scagnetto, I.: {pi-calculus in (Co)inductive-type
  theory}. Theor. Comput. Sci.  \textbf{253}(2),  239--285 (2001).
  \doi{10.1016/S0304-3975(00)00095-5}

\bibitem{JungSSSTBD15}
Jung, R., Swasey, D., Sieczkowski, F., Svendsen, K., Turon, A., Birkedal, L.,
  Dreyer, D.: Iris: Monoids and invariants as an orthogonal basis for
  concurrent reasoning. In: Rajamani, S.K., Walker, D. (eds.) Symposium on
  Principles of Programming Languages, {POPL} 2015. pp. 637--650. {ACM} (2015).
  \doi{10.1145/2676726.2676980}

\bibitem{K07}
Kobayashi, N.: {Type Systems for Concurrent Programs} (2007),
  \url{http://www.kb.ecei.tohoku.ac.jp/~koba/papers/tutorial-type-extended.pdf}

\bibitem{KPT96}
Kobayashi, N., Pierce, B.C., Turner, D.N.: {Linearity and the Pi-Calculus}. In:
  Symposium on Principles of Programming Languages, {POPL}. pp. 358--371. {ACM}
  Press (1996). \doi{10.1145/237721.237804}

\bibitem{Mackie}
Mackie, I.: {Lilac: A Functional Programming Language Based on Linear Logic}.
  J. Funct. Program.  \textbf{4}(4),  395--433 (1994).
  \doi{10.1017/S0956796800001131}

\bibitem{MatsakisK14}
Matsakis, N.D., II, F.S.K.: {The rust language}. In: High integrity language
  technology, {HILT}. pp. 103--104. {ACM} (2014). \doi{10.1145/2663171.2663188}

\bibitem{McBride2016}
McBride, C.: {I Got Plenty o' Nuttin'}. In: A List of Successes That Can Change
  the World, Lecture Notes in Computer Science, vol.~9600, pp. 207--233.
  Springer (2016). \doi{10.1007/978-3-319-30936-1\_12}

\bibitem{Milner99}
Milner, R.: {Communicating and mobile systems - the Pi-calculus}. Cambridge
  University Press (1999)

\bibitem{MilnerPW92}
Milner, R., Parrow, J., Walker, D.: {A Calculus of Mobile Processes, Parts I
  and II}. Inf. Comput.  \textbf{100}(1) (1992).
  \doi{10.1016/0890-5401(92)90008-4}

\bibitem{Orchard}
Orchard, D., Liepelt, V., III, H.E.: {Quantitative program reasoning with
  graded modal types}. Proc. {ACM} Program. Lang.  \textbf{3}({ICFP}),
  110:1--110:30 (2019). \doi{10.1145/3341714}

\bibitem{OrchardY16}
Orchard, D.A., Yoshida, N.: {Using session types as an effect system}. In: Gay,
  S., Alglave, J. (eds.) Programming Language Approaches to Concurrency- and
  Communication-cEntric Software, {PLACES} 2015. {EPTCS}, vol.~203, pp. 1--13
  (2015). \doi{10.4204/EPTCS.203.1}

\bibitem{Rouvoet2020}
Rouvoet, A., Poulsen, C.B., Krebbers, R., Visser, E.: {Intrinsically-typed
  definitional interpreters for linear, session-typed languages}. In: Certified
  Programs and Proofs, {CPP}. pp. 284--298. {ACM} (2020).
  \doi{10.1145/3372885.3373818}

\bibitem{Sangio01}
Sangiorgi, D., Walker, D.: {The Pi-Calculus - a theory of mobile processes}.
  Cambridge University Press (2001)

\bibitem{ScalasDHY17}
Scalas, A., Dardha, O., Hu, R., Yoshida, N.: {A Linear Decomposition of
  Multiparty Sessions for Safe Distributed Programming}. In: European
  Conference on Object-Oriented Programming, {ECOOP}. LIPIcs, vol.~74, pp.
  24:1--24:31. Schloss Dagstuhl - Leibniz-Zentrum f{\"{u}}r Informatik (2017).
  \doi{10.4230/LIPIcs.ECOOP.2017.24}

\bibitem{Thiemann2019}
Thiemann, P.: {Intrinsically-Typed Mechanized Semantics for Session Types} pp.
  19:1--19:15 (2019). \doi{10.1145/3354166.3354184}

\bibitem{TuronTABD13}
Turon, A.J., Thamsborg, J., Ahmed, A., Birkedal, L., Dreyer, D.: Logical
  relations for fine-grained concurrency. In: Giacobazzi, R., Cousot, R. (eds.)
  Symposium on Principles of Programming Languages, {POPL} '13. pp. 343--356.
  {ACM} (2013). \doi{10.1145/2429069.2429111}

\bibitem{Wadler90}
Wadler, P.: {Linear Types can Change the World!} In: Programming concepts and
  methods. p.~561. North-Holland (1990)

\bibitem{Zalakain2020Agda}
Zalakain, U., Dardha, O.: {Typing the Linear {{$\pi$}}-Calculus \textendash{}
  Formalisation in Agda}  (2021),
  \url{https://github.com/umazalakain/typing-linear-pi}

\end{thebibliography}

\clearpage
\appendix
\changetext{}{10em}{-5em}{-5em}{}

\section{From names to de Bruijn indices and back}
\label{from_to_deBruijn}

The syntax of the \picalc{} \cite{Sangio01} using channel names is given by the $\Raw$ grammar below:
\begin{mathpar}
  \datatype
  { }
  {\Raw : \Set}
  \; \textsc{Raw}
\end{mathpar}

\begin{equation*}
  \begin{aligned}
    \Raw ::= &\; \PO              &&\text{(inaction)}    \\ 
    |& \; (\new{\Name}) \; \Raw         &&\text{(restriction)} \\ 
    |& \; \comp{\Raw}{\Raw}       &&\text{(parallel)}    \\ 
    |& \; \recv{\Name}{\Name}\Raw &&\text{(input)}       \\ 
    |& \; \send{\Name}{\Name}\Raw &&\text{(output)}      \\
  \end{aligned}
\end{equation*}

Channel names and variables range over $x,y,z$ in $\Name$ and processes over $P,Q,R$ in $\Raw$.
Process $\PO$ denotes the terminated process, where no further communications can occur.
Process $(\new{}x)\; P$ creates a new channel $x$ bound with scope $P$.
Process $\comp{P}{Q}$ is the parallel composition of processes $P$ and $Q$.
Processes $\recv{x}{y} P$ and $\send{x}{y} P$ denote respectively, the input and output processes of a variable $y$ over a channel $x$, with continuation $P$.
Scope restriction $(\new{}x) \; P$ and input $\recv{x}{y} \; P$ are \emph{binders}, they are the only constructs that introduce bound names --- $x$ and $y$ in $P$, respectively.

In order to demonstrate the correspondence between a \picalc{} that uses names and one that uses de Bruijn indices, we provide conversion functions in both directions and prove that they are inverses of each other up to $\alpha$-conversion.

\paragraph*{From names to de Bruijn indices}
When we translate into de Bruijn indices we keep the original binder names around --- they will serve as name hints for when we translate back.
The translation function $\func{fromRaw}$ works recursively, keeping a context $ctx : \Names_n$ that maps the first $n$ indices to their names.
Named references within the process are substituted with their corresponding de Bruijn index.
We demand that the original process is well-scoped: that all its free variable names appear in $ctx$ --- this is decidable and we therefore automate the construction of such a proof term.
\begin{alignat*}{2}
    &\func{fromRaw} && : (ctx : \Names_n) \; (P : \Raw) \\
    &               && \to \WellScoped \; ctx \; P \to \Process_n
\end{alignat*}

\paragraph*{From de Bruijn indices to names}
The translation function $\func{toRaw}$ works recursively, keeping a context $ctx : \Names_n$ that maps the first $n$ indices to their names.
As some widely-used languages do, this translation function produces unique variable names.
These unique variable names use the naming scheme $<namehint>^{<n>}$, where $^{<n>}$ denotes that the name $<namehint>$ has already been bound $n$ times before.
\begin{equation*}
    \func{toRaw} : (ctx : \Names_n) \to \Process_n \to \Raw
\end{equation*}

\begin{example}[$\func{fromRaw}$ and $\func{toRaw}$]
  We illustrate the conversion functions from names to de Bruijn indices $(\func{fromRaw}$) and back ($\func{toRaw}$) with three processes $P,Q,R$  below.
  \begin{alignat*}{7}
    &P = (\new{x} ) && (\comp {\recv{x}{x} && \send{x}{z} && \PO} {(\new{y}) && (\send{x}{y} && \recv{y}{y} && \PO)}) \\
    &Q = \new{} && (\comp {\recv{0}{} && \send{0}{2} && \PO} {\new{} && (\send{1}{0} && \recv{0}{} && \PO)}) \\
    &R = (\new{x^0} ) && (\comp {\recv{x^0}{x^1} && \send{x^1}{z^0} && \PO} {(\new{y^0}) && (\send{x^0}{y^0} && \recv{y^0}{y^1} && \PO)})
  \end{alignat*}

  Process $P$ uses names $x,y,z$ and is translated via the conversion function $\func{fromRaw}$ into process $Q$, which uses de Bruijn indices.
  Process $Q$ is then translated via $\func{toRaw}$ into process $R$, which follows the \emph{Barendregt convention}\footnote{The Barendregt variable convention states that all bound variables/names in a process are distinct among each other and from the free variables/names.} and is $\alpha$-equivalent to the original process $P$.
\end{example}

In the following we present the main results that our conversion functions satisfy.

\begin{nilemma}
  Translating from de Bruijn indices to names via $\func{toRaw}$ results in a well-scoped process.
\end{nilemma}

\begin{nilemma}
  Translating from de Bruijn indices to names via $\func{toRaw}$ results in a process that follows the \emph{Barendregt convention}.
\end{nilemma}

\begin{nilemma}
  Translating from de Bruijn indices to names and back via $\func{fromRaw} \circ \func{toRaw}$ results in the same process modulo internal variable name hints.
\end{nilemma}

\begin{nilemma}
  Translating from names to de Bruijn indices and back via $\func{toRaw} \circ \func{fromRaw}$ results in the same process modulo $\alpha$-conversion.
\end{nilemma}

\begin{proof}
  All the above results are proved by induction on \textsc{Process}, \textsc{Var} (\autoref{def:syntax}) and \textsc{Raw}.
  Complete details can be found in our mechanisation in Agda in \cite{Zalakain2020Agda}.
\end{proof}

\section{Type Safety} \label{app:type-safety}

\paragraph*{Exchange}
This property states that the exchange of two variables preserves the well-typedness of a process.
We extend $\func{exchange}_i$  introduced in \autoref{def:equals} to exchange types in typing contexts and usage annotations in usage contexts.
\begin{nitheorem}[Exchange]
  \label{thm:exchange}
  Let $P$ be well typed in $\types{\gamma}{\Gamma}{P}{\Xi}$.
  Then, $\types{\func{exchange}_i \; \gamma}{\func{exchange}_i \; \Gamma}{\allowbreak \func{exchange}_i \; P}{\func{exchange}_i \; \Xi}$.
\end{nitheorem}

\begin{proof}
  All the above theorems are proved by induction on \textsc{Types} and \textsc{VarRef}.
  For details, refer to our mechanisation in Agda \cite{Zalakain2020Agda}.
\end{proof}

\paragraph*{Subject Congruence}
This property states that applying structural congruence (\autoref{def:equals}) to a well-typed process preserves its well-typedness.
To prove this result, we must first introduce lemmas that establish that certain syntactic manipulations can be inverted (\autoref{lm:lower-lift}, \autoref{lm:exchange-exchange}) and how unused variables relate to the preservation of leftovers (\autoref{lm:types-unused}).

\begin{nilemma}
  \label{lm:lower-lift}
  The function $\func{lower}_i \; P \; uP$ has an inverse $\func{lift}_i \; P$ that increments every $\textsc{Var}$ greater than or equal to $i$, such that $\func{lift}_i \; (\func{lower}_i \; P \; uP) \equiv P$.
\end{nilemma}
\begin{proof}
  By structural induction on \textsc{Process} and \textsc{Var}.
\end{proof}

\begin{nilemma}
  \label{lm:exchange-exchange}
  The function $\func{exchange}_i \; P$ is its own inverse: $\func{exchange}_i \; (\func{exchange}_i \; P) \equiv P$.
\end{nilemma}
\begin{proof}
  By structural induction on \textsc{Process} and \textsc{Var}.
\end{proof}

\begin{nilemma}
  \label{lm:types-unused}
  For all well-typed processes $\types{\gamma}{\Gamma}{P}{\Xi}$, if the variable $i$ is unused within $P$, then $\Gamma$ at $i$ is equal to $\Xi$ at $i$.
\end{nilemma}
\begin{proof}
  By induction on \textsc{Process} and \textsc{Var}.
\end{proof}

We are now in a position to prove subject congruence.

\begin{nitheorem}[Subject congruence]
  \label{thm:subject-congruence1}
  If $P \eq{} Q$ and $\types{\gamma}{\Gamma}{P}{\Xi}$, then $\types{\gamma}{\Gamma}{Q}{\Xi}$.
\end{nitheorem}

\begin{proof}
  The proof is by induction on \textsc{Equals} $\eq{}$.
  Here we only consider those cases that are not purely inductive: the base cases for $\constr{struct}$ and their symmetric variants.
  Full proof in \cite{Zalakain2020Agda}.
  We proceed by induction on \textsc{StructCong} $\eqeq$:
  \begin{itemize}
    \item
      Case $\constr{comp-assoc}$: trivial, as leftover typing is naturally associative.
    \item
      Case $\constr{comp-sym}$ for $\comp{P}{Q}$: we use framing (\autoref{thm:framing}) to shift the output context of $P$ to the one of $Q$; and the input context of $Q$ to the one of $P$.
    \item
      Case $\constr{comp-end}$: trivial, as the typing rule for $\PO$ has the same input and output contexts.
    \item
      Case $\constr{scope-end}$: we show that the usage annotation of the newly created channel must be $\lz$, making the proof trivial.
      In the opposite direction, we instantiate the newly created channel to a type $\unit$ and a usage annotation $\lz$.
    \item
      Case $\constr{scope-ext}$ for $\new (\comp{P}{Q})$: we need to show that $P$ preserves the usage annotations of the unused variable (\autoref{lm:types-unused}) and then use strengthening (\autoref{thm:strengthening}).
      In the reverse direction, we use weakening (\autoref{thm:weakening}) on $P$ and show that lowering and then lifting $P$ results in $P$ (\autoref{lm:lower-lift}).
    \item
      Case $\constr{scope-comm}$:
      we use exchange (\autoref{thm:exchange}), and for the reverse direction exchange and \autoref{lm:exchange-exchange} to show that exchanging two elements in $P$ twice leaves $P$ unchanged. \qed
  \end{itemize}
\end{proof}

\paragraph*{Substitution}
This result is key to proving subject reduction.
In \autoref{thm:substitution-generalization} we prove a generalised version of substitution, where the substitition $\subst{P}{j}{i}$ is on any variable $i$.
Then, in \autoref{thm:substitution} we instantiate the generalised version to the concrete case where $i$ is the most recently introduced variable $\constr{0}$, as required by subject reduction.

\begin{nitheorem}[Generalised substitution]
  \label{thm:substitution-generalization}
  Let process $P$ be well-typed in $\types{\gamma}{\Gamma_i}{P}{\Psi_i}$.
  The substituted variable at position $i$ can be split into $m$ in $\Gamma_i$, and into $n$ in $\Psi_i$.
  Substitution will take these usages $m$ and $n$ away from $i$ and transfer them to the variable $j$ we are substituting for.
  In other words, let there be some $\Gamma$, $\Psi$, $\Gamma_j$ and $\Psi_j$ such that:
  \begin{multicols}{2}
  \begin{itemize}
    \item $\contains{\gamma}{\Gamma_i}{i}{t}{m}{\Gamma}$
    \item $\contains{\gamma}{\Gamma_j}{j}{t}{m}{\Gamma}$
    \item $\contains{\gamma}{\Psi_i}{i}{t}{n}{\Psi}$
    \item $\contains{\gamma}{\Psi_j}{j}{t}{n}{\Psi}$
  \end{itemize}
  \end{multicols}
  Let $\Gamma$ and $\Psi$ be related such that $\opctx{\Gamma}{\Delta}{\Psi}$ for some $\Delta$.
  Let $\Delta$ have a usage annotation $\lz$ at position $i$, so that all consumption from $m$ to $n$ must happen in $P$.
  Then substituting $i$ to $j$ in $P$ will be well-typed in $\types{\gamma}{\Gamma_j}{\subst{P}{j}{i}}{\Psi_j}$.
\end{nitheorem}

\begin{proof}
  By induction on the derivation $\types{\gamma}{\Gamma_i}{P}{\Psi_i}$.
  \begin{itemize}
    \item
      For constructor $\PO$ we get $\Gamma_i \equiv \Psi_i$.
      From $\Delta_i \equiv \lz$ follows that $m \equiv n$.
      Therefore $\Gamma_j \equiv \Psi_j$ and $\constr{end}$ can be applied.

    \item
      For constructor $\new$ we proceed inductively, wrapping arrows $\ni_i m$, $\ni_j m$, $\ni_i n$ and $\ni_j n$ with $\suc$.
      
    \item
      For constructor $\recv{}{}$ we must split $\Delta$ to proceed inductively on the continuation.
      Observe that given the arrow from $\Gamma_i$ to $\Psi_i$ and given that $\Delta$ is $\lz$ at index $i$, there must exist some $\delta$ such that $\opsquared{m}{\delta}{n}$.
 l     \begin{itemize}
        \item
          If the input is on the variable being substituted, we split $m$ such that $\opsquared{m}{\li}{l}$ for some $l$, and construct an arrow $\containsusage{\Xi_i}{i}{l}{\Gamma}$ for the inductive call.
          Similarly, we construct for some $\Xi_j$ the arrows $\containsusage{\Gamma_j}{j}{\li}{\Xi_j}$ as the new input channel, and $\containsusage{\Xi_j}{j}{l}{\Gamma}$ for the inductive call.
        \item
          If the input is on a variable $x$ other than the one being substituted, we construct the arrows $\containsusage{\Xi_i}{i}{m}{\Theta}$ (for the inductive call) and $\containsusage{\Gamma}{x}{\li}{\Theta}$ for some $\Theta$.
          We then construct for some $\Xi_j$ the arrows $\containsusage{\Gamma_j}{x}{\li}{\Xi_j}$ (the new output channel) and $\containsusage{Xi_j}{j}{m}{\Theta}$ (for the inductive call).
          Given there exists a composition of arrows from $\Xi_i$ to $\Psi$, we conclude that $\Theta$ splits $\Delta$ such that $\opctx{\Gamma}{\Delta_1}{\Theta}$ and $\opctx{\Theta}{\Delta_2}{\Psi}$.
          As $\lz$ is a minimal element, then $\Delta_1$ must be $\lz$ at index $i$, and so must $\Delta_2$.
      \end{itemize}

    \item
      $\send{}{}$ applies the ideas outlined for the $\recv{}{}$ constructor to both the \textsc{VarRef} doing the output, and the \textsc{VarRef} for the sent data.

    \item
      For $\comp{}{}$ we first find a $\delta$, $\Theta$, $\Delta_1$ and $\Delta_2$ such that $\containsusage{\Xi_i}{i}{\delta}{\Theta}$ and $\opctx{\Gamma}{\Delta_1}{\Theta}$ and $\opctx{\Theta}{\Delta_2}{\Psi}$.
      Given $\Delta$ is $\lz$ at index $i$, we conclude that $\Delta_1$ and $\Delta_2$ are too.
      Observe that $\opsquared{m}{\delta}{\psi}$, where $\psi$ is the usage annotation at index $i$ consumed by the subprocess $P$.
      We construct an arrow $\containsusage{\Xi_j}{j}{\delta}{\Theta}$, for some $\Xi_j$.
      We can now make two inductive calls (on the derivation of $P$ and $Q$) and compose their results.

      \centering
      \begin{tikzpicture}
        \node (gamma-i) at (0,4)   {$\Gamma_i$};
        \node (gamma-m) at (0,2)   {$\Gamma$};
        \node (gamma-j) at (0,0)   {$\Gamma_j$};
        \node (xi-i)    at (3,3.5) {$\Xi_i$};
        \node (theta)   at (3,2.3) {$\Theta$};
        \node (delta-m) at (3,1.7) {};
        \node (xi-j)    at (3,0.5) {$\Xi_j$};
        \node (psi-i)   at (6,3)   {$\Psi_i$};
        \node (psi-m)   at (6,2)   {$\Psi$};
        \node (psi-j)   at (6,1)   {$\Psi_j$};

        \draw[-]  (gamma-m) -- (delta-m.center);
        \draw[->] (delta-m.center) -- node[align=center,below] {$\Delta$\\$\Delta_i = \lz$}(psi-m);

        \draw[->,densely dotted] (gamma-m) -- (theta);
        \draw[->,densely dotted] (theta) -- (psi-m);

        \draw[->] (gamma-i) -- node[left] {$\ni_i m$} (gamma-m);
        \draw[->] (gamma-j) -- node[left] {$\ni_j m$} (gamma-m);
        \draw[->] (psi-i) -- node[right] {$\ni_i n$} (psi-m);
        \draw[->] (psi-j) -- node[right] {$\ni_j n$} (psi-m);

        \draw[->] (gamma-i) -- node[above] {$\vdash P$} (xi-i);
        \draw[->] (xi-i) -- node[above] {$\vdash Q$} (psi-i);
        \draw[->,densely dotted] (gamma-j) -- node[below] {$\vdash \subst{P}{j}{i}$} (xi-j);
        \draw[->,densely dotted] (xi-j) -- node[below] {$\vdash \subst{Q}{j}{i}$} (psi-j);
        \draw[->,densely dotted] (xi-i) -- node[left] {$\ni_i l$} (theta);
        \draw[->,densely dotted] (xi-j) -- node[left] {$\ni_j l$} (theta);
      \end{tikzpicture}

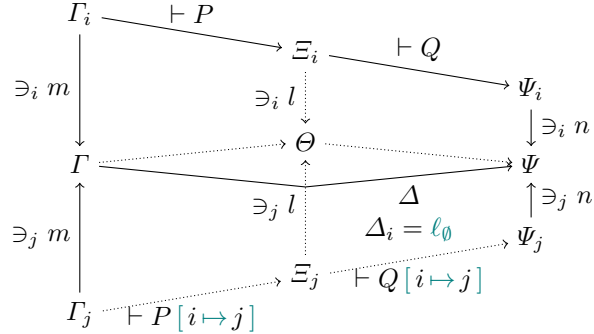
      Diagrammatic representation of the $\comp{}{}$ case for substitution.
      Continuous lines represent known facts, dotted lines proof obligations.

  \end{itemize}  
\end{proof}

\begin{nitheorem}[Substitution]
  \label{thm:substitution}
  Let process $P$ be well typed in $\types{\gamma \comma t}{\Gamma \comma m}{P}{\Psi \comma \lz}$.
  Let $\contains{\gamma}{\Psi}{j}{t}{m}{\Xi}$.
  Then, we can substitute the variable references to $\constr{0}$ in $P$ with $\suc j$ so that the result is well typed in $\types{\gamma \comma t}{\Gamma \comma m}{\subst{P}{\suc j}{\constr{0}}}{\Xi \comma m}$.
\end{nitheorem}
\begin{proof}
  For $\contains{\gamma}{\Gamma}{j}{t}{m}{\Theta}$ and $\types{\gamma \comma t}{\Theta \comma m}{P}{\Xi \comma \lz}$ for some $\Theta$, we use framing to derive them.
  Then, we use these to apply \autoref{thm:substitution-generalization}.
\end{proof}

\paragraph*{Subject Reduction}
Finally we are ready to present our main result, stating  that if $P$ is well typed and it reduces to $Q$, then $Q$ is well typed.
The relation between the typing contexts used to type $P$ and $Q$ will be explained in \autoref{thm:subject-reduction1}.
In the \picalc{} we distinguish between a reduction $P \reduce{\constr{internal}} Q$ on a channel internal to $P$, and a reduction $P \reduce{\constr{external} \; i} Q$ on a channel $i$ external to $P$ (refer to \autoref{semantics}).
We first introduce an auxiliary lemma:

\begin{nilemma}
  \label{lm:comm-capable}
  Every input usage context $\Gamma$ of a well-typed process $\types{\gamma}{\Gamma}{P}{\Delta}$ that reduces by communicating on a channel external (that is, $P \reduce{\constr{external} \; i} Q$ for some $Q$) has a multiplicity of at least $\lio$ at index $i$.
\end{nilemma}

\begin{proof}
  By induction on the reduction derivation $P \reduce{\constr{external \; i}}Q$.
\end{proof}

\begin{nitheorem}[Subject reduction]
  \label{thm:subject-reduction1}
  Let $P$ be well typed in $\types{\gamma}{\Gamma}{P}{\Xi}$ and reduce such that $P \reduce{c} Q$.
  \begin{itemize}
    \item If $c$ is $\constr{internal}$, then $\types{\gamma}{\Gamma}{Q}{\Xi}$.
    \item If $c$ is $\constr{external} \; i$ and $\containsusage{\Gamma}{i}{\lio}{\Delta}$, then $\types{\gamma}{\Delta}{Q}{\Xi}$.
  \end{itemize}
\end{nitheorem}

\begin{proof}
  By induction on $P \reduce{c} Q$. For the full details refer to our mechanisation in Agda.  
  \begin{itemize}
    \item
    Case $\constr{comm}$: we apply framing (\autoref{thm:framing}) (to rearrange the assumptions), substitution (\autoref{thm:substitution}) and strengthening (\autoref{thm:strengthening}).
  
    \item
    Case $\constr{par}$: by induction on the process that is being reduced.

    \item
    Case $\constr{res}$: case split on channel $c$:
    if $\constr{internal}$ proceed inductively;
    if $\constr{external}\; \constr{0}$ (i.e. the channel introduced by scope restriction) use \autoref{lm:comm-capable} to subtract $\lio$ from the channel's usage annotation and proceed inductively;
    if $\constr{external}\; (\suc i)$ proceed inductively.

    \item
    Case $\constr{struct}$: we apply subject congruence (\autoref{thm:subject-congruence1}) and proceed inductively. \qed
  \end{itemize}
\end{proof}

\end{document}